\documentclass[11pt]{iopart}
\usepackage{epic}
\usepackage{epsfig}
\usepackage{subfigure}
\usepackage{curves}
\usepackage{letterspace}
\usepackage{rotating}
\begin{document}

\newcommand{\abs}[1]{\mid#1\mid}
\newcommand{\DoT}[1]{\begin{turn}{-180}\raisebox{-1.7ex}{#1}\end{turn}}

\title{Neutrino masses and the structure of the weak gauge boson}
\author{V.N.Yershov}
\address{ University College London, Mullard Space Science Laboratory,\\
 Holmbury St.Mary, Dorking RH5 6NT, United Kingdom}
\ead{vny@mssl.ucl.ac.uk}

\date{January 15, 2003}

 \begin{abstract}
It is supposed that the electron neutrino mass is
related to the structures and masses of the $W^\pm$
and $Z^0$ bosons. Using a composite model of
fermions (described elsewhere), it is shown that the 
massless neutrino is not consistent with the high
values of the experimental masses of $W^\pm$ and $Z^0$.
Consistency can be achieved on the assumption that
the electron-neutrino has a mass of about 4.5 meV.
Masses of the muon- and tau-neutrinos are also estimated.     
\end{abstract}


\section{Introduction}

The results of atmospheric and solar neutrino 
observations~\cite{fukuda98} indicate that neutrinos have
non-zero mass. Although very small, the neutrino mass 
can play a key role in new physics underlying the
Standard Model of particle physics \cite{akhmedov00}.
The Standard Model considers all three neutrino flavor states ($\nu_e$,
$\nu_\mu$, and $\nu_\tau$) as massless. Thus, observations
show the Standard Model needs some corrections or extensions.

So far, many models have been proposed for the massive 
neutrino~\cite{pascoli01}, and many experiments have been
carried out for measuring this  mass. Experiments with
tritium $\beta$-decay gave controversial results: the  neutrino
mass squared appears to be negative.~\cite{ingraham00}
The current experimental upper limits for the 
neutrino masses (by the Particle Data
Group) \cite{properties}:

\hspace{1.0cm} $m_{\nu_e} \le 3\cdot 10^{-6}$, \hspace{0.5cm}
$m_{\nu_\mu} \le 0.19$, \hspace{0.5cm}
$m_{\nu_\tau} \le 18.2$ \hspace{0.5cm} (in $MeV/c^2$) \\
are based on the averaged data from kinematic 
determinations. Astrophysical data~\cite{elgar02}
give stronger constraints:

 \hspace{1cm}$\sum_i{m_i} < 1.8$ \hspace{0.2cm} $eV$ \hspace{0.7cm} (for all species). \\
Recently, the author has proposed  a new composite model
of fermions \cite{yersh02}, in which the neutrino is described
as a Majorana particle, almost massless in free states, but
acquiring a substantial mass if bonded to a charged particle.
The model satisfactorily (to an accuracy of $10^{-6}$) predicts
masses of the fundamental fermions (quarks and charged leptons) 
but, being approximate, it cannot reproduce masses of 
neutral particles, such as neutrinos or photons, as well as
of the weak gauge bosons ($Z^0$ and $W^\pm$).
Here we discuss this problem and propose its possible solution 
in the form of a hypothetical symmetry between fermion and boson
mass structures.

\section{Hypothetical structure of $W^\pm$}

The model~\cite{yersh02} is based on the idea of compositeness 
of fermions. Properties of their primitive constituents
(``preons'') are limited to the SU(3)-symmetry of colored-charges
 and a mass. For simplicity, we use unit values
for the preon's mass and charge.
It is supposed that by rotational transformations the preon 
can be translated into one of its six possible color/charge states.
Preons of different colors and charges should group in equilibrium
configurations corresponding to the symmetry of their potentials.
These configurations are identified with known particles. It was
shown that masses of the composite fermions
can be calculated as sums of the masses
of their constituents (preons), if the composite
particle is ``rigid''. For the ``non-rigid'' structures,
the following formula
\begin{equation}
\label{eq.mtotal}
m=(\sum_{i=1}^N{m_i})(\sum_{i=1}^N{1/m_i})^{-1} .
\end{equation}
can be used. The ``rigidness'' of a particle is determined by 
the structure of its effective potential.
Simpler potentials (having, say, a single minimum)
should correspond to more ``rigid'' structures. These are  -- in our
model -- the first family fermions. The second and third families 
are clusters of particles rather than rigid structures.  
The masses in (\ref{eq.mtotal}) are expressed in units of the preon's
mass. They can be converted into
conventional units, say,  proton mass units ($m_p$), by calculating $m_p$
with (\ref{eq.mtotal}) and dividing all other calculated masses by 
this value. 
The masses can be converted also into $MeV/c^2$ by using
a conversion factor $k_\Pi=0.056777673 \hspace{0.2cm} [MeV/\Pi]$ 
derived from the experimental fermion masses~\cite{yersh02}
($\Pi$ stands for the preon). 
  
Three like-charged preons with complementary colors will combine 
in a {\sf Y}-shaped triplet. The triplet has a mass of three
preons, $m_{\sf Y}=3$, and a charge $q_{\sf Y}=\pm 3$ preon units.
The minimal-length closed loop composed of six preon-antipreon
triplet pairs ${\sf Y} \overline{\sf Y}$, can be associated with 
the electron-neutrino, $\nu_e=6({\sf Y} \overline{\sf Y})$.  
The charges and masses of the adjacent unlike-charged constituents 
(${\sf Y}$ and  $\overline{\sf Y}$) are considered to almost 
cancel mutually.
In this notation, the structure of the electron/positron can be
written as $e^-=3{\sf Y}$ / $e^+=3\overline{\sf Y}$,
with a charge $q_e=\pm 9$ and a mass $m_e=9$ preon units
($9\times k_\Pi=0.511$ $MeV$).
In our model, a neutral massless particle, if ``rigidly''
coupled to a charged particle, restores its mass.
The mass of such a system will be equal to the sum 
of the masses of its preons, and its charge will correspond to the 
charge of its charged component. 
 For instance, the mass of the system
 ${\sf Y}6({\sf Y} \overline{\sf Y})={\sf Y}\nu_e$ is the sum
 of $m_{{\sf Y}}=3$ and 36 unit masses of the
 preons constituing $\nu_e$  (total 39 preon mass units). 
Its charge ($q_{{\sf Y}6({\sf Y} \overline{\sf Y})}=-3$)
corresponds to the charge of a single {\sf Y}-particle.
We identify the chain $(\nu_e\overline{{\sf Y}}) 
\nu_e (\overline{{\sf Y}}\nu_e)$ with the $up$-quark.
Its charge is equal to the charge of two $\overline{\sf Y}$-particles
($+6$) and its mass is the sum of two 
$m_{{\sf Y}6({\sf Y} \overline{\sf Y})}$:
total 78 mass units or $78\times k_\Pi=4.42$ $MeV$. The 
positively charged $up$-quark, combined with the negatively
charged structure $3{\sf Y}6({\sf Y}\overline{\sf Y})=e\nu_e$,
forms the $down$-quark,
$3{\sf Y}6({\sf Y}\overline{\sf Y})(\nu_e\overline{{\sf Y}}) 
\nu_e (\overline{{\sf Y}}\nu_e)$,  with its charge of $-9+3+3=-3$ units
and a mass of $9+36+78=123$ units ($123\times k_\Pi=6.98$ $MeV$).
The structure $W^-=3{\sf Y}6({\sf Y}\overline{\sf Y})$, which has 
properties of the weak gauge boson (see Table 2 in 
Ref.~\cite{yersh02}), consists of 45 preons (9 preons of the electron
and 36 preons of $\nu_e$). 
Thus, its mass  should be about 2.6 $MeV$ if it is in a  
rigid structure, or almost zero if considered as 
a cluster of $\nu_e$ and $e$. 
However, the experimental mass of $W^\pm$ is $80~GeV$, indicating
that (\ref{eq.mtotal}) is not applicable to bosons. Even for 
$m_{\nu_e} \neq 0$ the calculated mass of $W^\pm$
increases very little, still disagreeing with experiment.
The problem can be resolved by using a reciprocal formula 
\begin{equation}
\label{eq.mtotal2}
m=(\sum_{i=1}^N{m_i})^{-1}(\sum_{i=1}^N{1/m_i})
\end{equation}
for bosons, instead of (\ref{eq.mtotal}), assuming a symmetrical
relationship between the boson and fermion mass structures. The physical 
justification for this assumption is analogous to that of supersymmetry,
but here it is based on the dualism of space introduced
in Ref.~\cite{yersh02}.
We suppose that two reciprocal manifestations of space
are interchangeable for fermions and bosons. Eq.~(\ref{eq.mtotal2})
reproduces the large $W^\pm$ and $Z^0$ masses. In fact,
had the electron-neutrino mass be taken as zero, the masses of $W^\pm$ and $Z^0$ 
would become infinite. Varying $m_{\nu_e}$
in (\ref{eq.mtotal2}), one can obtain a realistic mass for
 the weak gauge boson. 
The following electron-neutrino mass
\begin{equation}
\label{eq.nue}
m_{\nu_e}=(4.453\pm 0.002)\cdot 10^{-3} \hspace{0.5cm}[eV]
\end{equation}
(or $7.864\cdot 10^{-8}$ preon mass units) fits 
the experimental value~\cite{properties} of $m^{exp}_W=80.43\pm 0.04~GeV$
(1416578 preon mass units) and the mass of the electron,
$m_e=9$, expressed in preon mass units. The tolerance interval 
is translated from $m^{exp}_W$.

\section{The neutrinos $\nu_\mu$ and $\nu_\tau$}

In \cite{yersh02} the muon- and tau-neutrinos 
are formed  by consecutive additions of the
unlike-charged pairs $(\nu_e {\sf Y})$ and $(\nu_e \overline{{\sf Y}})$ to 
the electron-neutrino. We conjecture that the muon-neutrino have one
such pair in its structure:
\begin{equation}
\label{eq.numu}
\nu_\mu=(\nu_e {\sf Y}) \nu_e (\overline{{\sf Y}}\nu_e),
\end{equation}    
and the tau-neutrino might have two such pairs:
\begin{equation}
\label{eq.nutau}
\nu_\tau=(\nu_e \overline{{\sf Y}}) \nu_e (\nu_e {\sf Y}) \nu_e
 (\overline{{\sf Y}}\nu_e) \nu_e ({\sf Y}\nu_e).
\end{equation}
Both $\nu_\mu$ and $\nu_\tau$ are constituents 
of the tau-lepton where, together with $\mu$, they are 
clustered in two components, $\mu \nu_\mu$ and
 $\nu_\tau$: 
\begin{equation}
\label{eq.tau}
\raisebox{1.1cm}{$\tau^-=$\hspace{0.1cm}}{
\stackrel{
           \underbrace{{\raisebox{0.6cm}{[}}\stackrel{{\raisebox{0.1cm}{$\mu^-$}}}
                                {\overbrace{
                                            {\stackrel{{\raisebox{-0.1cm}{$\underbrace{\nu_e e^-}$}}}{{\raisebox{-0.1cm}{$W^-$}}}} 
                                            \stackrel{\underbrace{(\overline{{\sf Y}}\nu_e) \nu_e (\nu_e {\sf Y})}}{\nu_\mu}
                                           }
                                } 
                      \hspace{0.2cm} \stackrel{{\raisebox{0.1cm}{$\nu_\mu$}}}
                                              {{\raisebox{0.6cm}{{$\overbrace{
                                                          (\overline{{\sf Y}}\nu_e) \nu_e (\nu_e {\sf Y})
                                                         }$
                                              }]}}}}            
         }
         {\raisebox{-0.2cm}{\fbox{{\tiny 1}}}}
       } 
{
\stackrel{\raisebox{-0.7cm}{{   
           $\underbrace{[\stackrel{\raisebox{0.1cm}{{$\nu_\tau$}}}
                             {\overbrace{
                                         (\overline{{\sf Y}}\nu_e) \nu_e (\nu_e {\sf Y}) \nu_e (\overline{{\sf Y}}\nu_e) 
                                                   \nu_e (\nu_e{\sf Y})                                           
                              }          }
                      ]}                     
         $}}}
         {\raisebox{0.4cm}{\fbox{{\tiny 2}}}}
}\raisebox{1.1cm}{\hspace{0.1cm}.} 
\end{equation} 
Here the clustered components are grouped in
brackets and marked with the boxed numbers. 
The number of preons (mass) of the first component is
\begin{equation}
\label{eq.nn1}
 m_1=n_\mu+n_{\nu_\mu}=n_W+n_{\nu_\mu}+n_{\nu_\mu}=45+78+78=201.
\end{equation}
With this value in (\ref{eq.mtotal2}) and taking into account
 the known experimental mass~\cite{properties} of the tau-lepton,
$m_2=m_\tau^{exp}=1776.99^{+0.29}_{-0.26}$  $MeV$  
(31297.3 in preon mass units), one can translate $m_\tau^{exp}$ 
into the tau-neutrino mass:
\begin{equation}
\label{eq.mnutau}
m_{\nu_\tau}=(9.0253^{+0.0011}_{-0.0018})\cdot 10^{-3} \hspace{0.2cm} [eV]
\end{equation}
(or $1.5896^{+0.0002}_{-0.0003}\cdot 10^{-7}$ preon mass units).

The muon-neutrino mass cannot be estimated in the same way as $m_{\nu_e}$ 
or $m_{\nu_\tau}$, because there are no particles containing
$\nu_\mu$ as a separate component within their structures.
For example, the muon structure (see Ref.~\cite{yersh02} for details)
can be written as:
\begin{equation}
\label{eq.muon}
\raisebox{0.1cm}{$\mu^-=$\hspace{0.1cm}}
\stackrel{
          [\stackrel{{\raisebox{0.1cm}{$W^-$}}}
                    {\overbrace{
                                \nu_e e^- 
                               }
                    }
         }
         {\raisebox{-0.4cm}{\hspace{0.3cm} \fbox{\tiny 1}}}
\stackrel{
          \stackrel{{\raisebox{0.1cm}{$\nu_\mu$}}}
                   { \hspace{0.1cm}
                     \overbrace{
                                 \overline{{\sf Y}}]  
                                     [\nu_e \nu_e (\nu_e {\sf Y})],
                               }
                   }
         }
         {\raisebox{-0.4cm}{\hspace{0.6cm} \fbox{\tiny 2}}} 
\raisebox{0.1cm}{\hspace{0.1cm}} 
\end{equation} 
which includes $\nu_\mu$. But, having its own substructure,
$\nu_\mu$ is shared here between
two components, \fbox{\tiny 1} and \fbox{\tiny 2}. Using the known
experimental muon mass \cite{properties},
$m_\mu^{exp}=105.658$ [$MeV$] (1860.9168 preon mass units)
and $m_W=45$ preon mass units, one can evaluate the upper limit for
the muon-neutrino mass: $m_{\nu_\mu} \leq 0.68$ $eV$. However, it
would  be more logical to suppose that $m_{\nu_\mu}$ lies
somewhere between
$m_{\nu_e}=4.5~meV$ and
 $m_{\nu_\tau}=9.0~meV$; that is, 
\begin{equation}
\label{eq.mnumu}
m_{\nu_\mu} \approx 6 \cdot 10^{-3} \hspace{0.2cm} [eV].
\end{equation}

\section{Structure of $Z^0$}

The $Z^0$-boson is a photon-like particle with high mass. 
According to Ref.~\cite{yersh02}, its structure
can be written as:
\begin{equation}
\label{eq.zet}
\raisebox{0.1cm}{$Z^0=$\hspace{0.1cm}}
   \stackrel{W^-}
            {\overbrace{
                        \nu_e e^- 
                       }
            }
   \stackrel{W^+}
            {\overbrace{
                        e^+ \nu_e
                       }
            }
\raisebox{0.1cm}{\hspace{0.1cm}.} 
\end{equation} 
From this, by using $m_{e^-}=m_{e^+}=9$ ($m_{e^-e^+}=18$ 
preon mass units) and the known experimental
mass of $Z^0$ ($m^{exp}_{z^\circ}=91.1876~GeV$ or 1606047 preon mass units), 
one can estimate the electron-neutrino
mass: $3.928\cdot 10^{-3}~eV$ (or $6.9183\cdot 10^{-8}$ preon mass
units). This is smaller than (\ref{eq.nue}), but the 
structure of $Z^0$ is more complex than that of $W^\pm$. Thus, 
the possible underestimation of $m_{\nu_e}$
might be caused by overestimation of $m_{e^-e^+}$.   
Using (\ref{eq.nue}) and $m^{exp}_{z^\circ}=91.1876\pm 0.0021~GeV$,
one can calculate
\begin{equation}
\label{eq.mee}
m_{e^+e^-}=0.90140\pm 0.00004 \hspace{0.2cm} [MeV]
\end{equation}
(or $15.876\pm 0.008$ preon mass units).

\section{Conclusions}

The  symmetry introduced here between bosons and fermions 
permits the estimation of the neutrino masses, based on the
experimental values for  
$m_{{\rm W}^{\pm}}$, $m_{\tau}$, or $m_{z^0}$.
However, our estimations are made for the neutrinos
interacting with other particles. Thus, the calculated  
$m_{\nu_e}\approx 4.5~meV$, $m_{\nu_\mu}\approx 6~meV$,
and $m_{\nu_\tau}\approx 9~meV$ should probably be considered
only as upper limits for the free neutrino masses. 


\end{document}